\documentclass[a4paper,11pt]{article}
\usepackage[utf8]{inputenc}
\usepackage[longnamesfirst]{natbib}

\usepackage[top=38truemm,bottom=38truemm,left=28truemm,right=28truemm]{geometry}
\usepackage{physics}
\usepackage{url}

\usepackage{mathtools}
\mathtoolsset{showonlyrefs=true} 

\usepackage{amsthm, amssymb, amsmath, appendix}

\theoremstyle{definition}
\newtheorem{theorem}{Theorem}
\newtheorem{lem}{Lemma}

\newtheorem{proposition}{Proposition}

\newtheorem{claim}{Claim}

\newcommand{\argmax}{\mathop{\rm arg ~ max}\limits}

\usepackage{tikz}
\usetikzlibrary{intersections, calc, arrows.meta}

\makeatletter

\@addtoreset{figure}{section}
\@addtoreset{table}{section}
\makeatother 

\title{\huge{Weak independence of irrelevant alternatives and \\ generalized Nash bargaining solutions}\thanks{This paper is based on Chapter 2 of the author's Master's Thesis submitted to Hitotsubashi University. The author is grateful to Walter Bossert, Noriaki Kiguchi, Koichi Tadenuma, Norio Takeoka, Tsubasa Yamashita for their helpful comments.}}
\author{Kensei Nakamura\thanks{Graduate School of Economics, Hitotsubashi University, Kunitachi, Tokyo 186-8601, Japan. E-mail: kensei.nakamura.econ@gmail.com (ORCID: 0009-0008-4549-4215)}}
\date{This version: \today}

\begin{document}

\maketitle

\vspace{10mm}

\begin{abstract}
In Nash's (\citeyear{nash1950bargaining}) seminal result, independence of irrelevant alternatives (IIA) plays a central role, but it has long been a subject of criticism in axiomatic bargaining theory. 
This paper examines the implication of a weak version of IIA in multi-valued bargaining solutions defined on non-convex bargaining problems. 
We show that if a solution satisfies weak IIA together with standard axioms, it can be represented, like the Nash solution, using weighted products of normalized utility levels. 
In this representation, the weight assigned to players for evaluating each agreement is determined endogenously through a two-stage optimization process. 
These solutions bridge the two dominant solution concepts, the Nash solution and the Kalai-Smorodinsky solution (\citealp{kalai1975other}). 
Furthermore, we consider special cases of these solutions in the context of bargaining over linear production technologies. 
\vspace{5mm}
\end{abstract}

\newpage
\section{Introduction}
\label{sec_intro}

In axiomatic bargaining theory, Nash's (\citeyear{nash1950bargaining}) independence of irrelevant alternatives (IIA) has long been a subject of criticism. This axiom requires that if an outcome chosen in the original problem remains feasible in a smaller problem, then it should remain to be selected in that  problem. Since this axiom postulates consistency in choices even when the balance of bargaining power among players changes significantly due to shrinking of the feasible set, it prevents bargaining solutions from adequately reflecting shifts in bargaining power. To overcome this limitation, many researchers have proposed alternative axiomatizations of the Nash bargaining solution without using IIA, as well as other alternative bargaining solutions such as the Kalai-Smorodinsky bargaining solution (\citealp{kalai1975other}).\footnote{For a series of comprehensive surveys, see \cite{thomson1981nash,Thomson1994CHAP,Thomson2022RED}.}

This paper considers multi-valued bargaining solutions defined on the non-convex bargaining problems and examines a general class of bargaining solutions that satisfy a weakened version of Nash's IIA together with the basic axioms shared by many well-known bargaining solutions.\footnote{Note that the standard convexity assumption is justified by the randomization of outcomes.
To examine situations where the randomization is not valid or prohibited, we include non-convex problems into the domain. Many papers also considered the bargaining problem without the convexity assumption
(e.g., \citealp{conley1991bargaining,herrero1989nash,mariotti2000maximal,ok1999revealed,xu2006alternative,zhou1997nash}).
For surveys of non-convex problems, see Section 4 of \citet{Thomson2022RED} and \citet{xu2020nonconvex}.  
} This weakened IIA imposes consistency in choices only when the maximum utility levels that players can achieve are identical among players in both the original and the new bargaining problems. By introducing this restriction, the chosen outcomes can change in response to shifts in bargaining power.

We show that bargaining solutions satisfying these axioms can be represented using  weighted products of utility values normalized by each player's maximum utility level, similar to the Nash bargaining solution.\footnote{In many papers, the Nash bargaining solution is defined as the maximizer of the products of utility levels. However, since it satisfies the scale-invariance property, it can be rewritten as the maximizer of the products of normalized utility levels. For a formal definition, see Section \ref{sec2_brgainingProblem}.} While the Nash bargaining solution directly uses the product of normalized utility levels to evaluate each outcome, the solutions characterized in this paper endogenously calculate ``optimal" weights over players for each outcome. This process can be interpreted as a decision-making rule of an arbitrator who consults multiple experts before making a decision: Each inequality-averse expert suggests their evaluation of an outcome, and then the optimistic arbitrator chooses the maximum value from  the suggestions. 

This class of bargaining solutions includes the Nash bargaining solution as an extreme case. Moreover, the Kalai-Smorodinsky bargaining solution, which has been widely analyzed as an alternative to the Nash solution, is included as another extreme case in this class. (It is worth noting that several studies have characterized the Kalai-Smorodinsky solution and related bargaining solutions using axioms similar to our weakened IIA; see, for example, \cite{DUBRA2001131} and \citet{xu2006alternative}.) Thus, our results can be seen as characterizing a general class of bargaining solutions that bridges the two most widely known bargaining solutions.

In research on decision-making models under uncertainty, preferences represented by two-stage optimizations of opposite direction interests have been examined (\citealp{lehrer2011justifiable,chandrasekher2022dual}).  
Mathematically, this paper is closely related to those works. \citet{chandrasekher2022dual} showed that if an objective function is positively homogeneous and constant-additive, then it can be represented as a preference of a dual-self individual, that is, an evaluation rule with a two-stage optimization problem. Moreover, they argued that if a function is only constant-additive, then the above result can be generalized and the function can still be represented as a preference of a dual-self decision maker. 
In both results, the evaluation rules for uncertain prospects are based on expected utility values (i.e., weighted sums of utility levels). 
This paper extends these results by showing that if an objective function defined on the nonnegative orthant is positively homogeneous, it can also be written as a two-stage optimization problem. However, in this case, the calculation process is not based on the weighted sums but instead on the Nash products.

Furthermore, we examine bargaining over linear production technologies.  By introducing additional axioms compelling in this context, we specify simpler and more understandable subclasses of bargaining solutions. 
These subclasses of solutions are also highly related to results in decision theory such as \citet{gilboa1989maxmin}. 
Thus, this paper can be considered as providing guidance on how to connect two distinct areas of literature: axiomatic bargaining theory and decision theory.

Our weakened version of IIA is convincing when it is desirable to reflect players' bargaining power. However, if an exogenously given welfare function exists, the criticism of IIA loses its validity. Rather, the scale-invariance axiom becomes questionable in Nash's theorem since this axiom postulates that solutions should be responsive to changes in problems. 
We show that a similar result to our main theorem can be obtained in these cases. 
Specifically, our result shows that if the weakened IIA and the scale-invariance axiom are replaced with the full IIA and the axiom of homogeneity, then the solutions can still be represented using weighted products of the utility levels, but without normalization.


 This paper is organized as follows: Section \ref{sec2_brgainingProblem} introduces the bargaining problem and bargaining solutions. Section \ref{sec2_axioms} presents axioms for bargaining solutions. Section \ref{sec2_main} provides our main characterization results. 
In Section \ref{sec2_tech}, we consider bargaining over linear production technologies and provide two new characterization results. Section \ref{sec2_homo} examines bargaining solutions that satisfy full IIA and homogeneity instead of the scale-invariance property. 
Section \ref{sec2_concl} provides concluding remarks. 
All Proofs are in Appendix. 

\section{The bargaining problem}
\label{sec2_brgainingProblem}

\subsection{Preliminary}

Let $N = \{1,2, \cdots, n \} $ be set of players with $n\geq 2$. 
They stay at the origin $(0,0, \cdots 0) \in \mathbb{R}^n$, denoted by $\mathbf{0}$, if they fail to reach an agreement.\footnote{
Let $\mathbb{R}$ (resp. $\mathbb{R}_{+}, ~ \mathbb{R}_{++}$) denote the set of real numbers (resp. nonnegative numbers, positive numbers). Let $\mathbb{R}^n$ (resp. $\mathbb{R}^n_{+}, ~ \mathbb{R}^n_{++}$) denote the $n$-fold Cartesian product of $\mathbb{R}$ (resp. $\mathbb{R}_{+}, ~ \mathbb{R}_{++} $). Let $\mathbb{N}$ be the set of natural numbers.
}
A \textit{(bargaining) problem} is a nonempty set $S \subset \mathbb{R}^n_{+}$ of utility vectors the players can achieve through unanimous agreement. 
We assume that problems $S$ satisfy the following conditions:  
\begin{itemize}
    \item The set $S$ is  compact and comprehensive (i.e., for all $x, y \in \mathbb{R}^n_+$, if $x \in S$ and $y\leq x $, then $y\in  S$).\footnote{We write $a \gg b$ if $a_i > b_i$ for all $i\in N$, and $a \geq b$ if $a_i \geq b_i$ for all $i\in N$. We define $\ll$ and $\leq$ in the same way. }  
    \item For all $i\in N$, there exists $x\in S$ such that $x_i > 0$. 
\end{itemize}
Note that we do not impose convexity on the problems.
The set of all problems is denoted by $\mathcal{B}$. 

For all $S\in \mathcal{B}$, define $b(S) \in \mathbb{R}^n_{++}$ as $b_i (S) = \max_{x\in X} x_i$ for each $i\in N$.  Let $\mathcal{B}^\text{Eq}$ be the set of problems where the maximum utility levels that  players can achieve are equal, that is,  $\mathcal{B}^\text{Eq} = \{ S\in \mathcal{B} \mid b_i (S) = b_j (S) ~~\text{for all} ~~ i,j\in N \}$.  

Let $\mathbf{1}$ denote $(1 , 1, \cdots, 1) \in \mathbb{R}^n$. 
A function $\pi: N\rightarrow N$ is a \textit{permutation} if it is a one-to-one function. Let $\Pi$ be the set of permutations. 
For all $x\in \mathbb{R}^n$, let $x^\pi = (x_{\pi (1)}, x_{\pi (2)}, \cdots, x_{\pi (n)})$. 
We say that a set $A\subset \mathbb{R}^n$ is \textit{symmetric} if  for all  $\pi \in \Pi$, $A =\{ x^\pi \mid x\in A\}$. 
For all $x\in \mathbb{R}^n$, $(x_{(1)}, x_{(2)}, \cdots, x_{(n)})$ denotes a rearrangement of $x$ such that $x_{(1)}\leq  x_{(2)}\leq  \cdots\leq x_{(n)}$, with the ties being broken arbitrarily. 
For all $a,x  \in \mathbb{R}^n$, let $ax = (a_1 x_1, a_2 x_2, \cdots a_n x_n)$. For all $a\in \mathbb{R}^n$ and $S\subset \mathbb{R}^n$, let $aS = \{ ax \mid x\in S \}$.

For a set $A \subset \mathbb{R}^n$, let $\text{cmp} \,  A$ denote the comprehensive full of $A$. That is, $\text{cmp} \,A$ is the smallest comprehensive set including $A$ and can be written as 
\begin{equation*}
    \text{cmp} \,A = \{ x\in \mathbb{R}^n_{+} \mid \text{there exists $a\in A$ such that $a\geq x$} \}.
\end{equation*}
Similarly, for a set $A \subset \mathbb{R}^n$, the symmetric comprehensive hull of $A$ is denoted by $\text{scmp} \,A$, that is,
\begin{equation*}
    \text{scmp}\, A = \{ x\in \mathbb{R}^n_+ \mid \text{there exist $a\in A$ and  $\pi \in \Pi$ such that $a^\pi \geq x$} \}.
\end{equation*}

\subsection{Bargaining solutions}

A \textit{(bargaining) solution} $F$ assigns a nonempty subset $F(S)$ of $S$ for each bargaining problem $S\in \mathcal{B}$.
The solution introduced in \citet{nash1950bargaining}, which we call the \textit{Nash solution}, is the most well-known solution. 
Formally, the Nash solution $F^\text{Nash}$ is defined as for all $S \in \mathcal{B}$, 
\begin{equation*}
     F^\text{Nash}(S) = \argmax_{x\in S} \prod_{i\in N} x_i. 
\end{equation*}
Note that the Nash solution can be rewritten as follows: For all $S\in \mathcal{B}$, 
\begin{equation*}
    F^\text{Nash} (S) = \argmax_{x\in S} \prod_{i\in N} \qty({x_i \over b_i (S)}). 
\end{equation*}

The \textit{Kalai-Smorodinsky solution} $F^\text{KS}$, defined as for all $S\in \mathcal{B}$
\begin{equation*}
    F^\text{KS} (S) = \argmax_{x\in S} \min_{i\in N} {x_i \over b_i (S)}, 
\end{equation*}
is also widely considered. 
The above  definition differs from the original one, but they are equivalent in the convex problems.

\section{Axioms for solutions}
\label{sec2_axioms}

Natural or reasonable properties for solutions, referred to as \textit{axioms}, have been examined in the literature. 
We start with an axiom of efficiency. This axiom  requires that for all problems, all chosen outcomes should be weakly Pareto efficient and at least one of them should be strictly Pareto efficient.
\begin{description}
    \item[\bf Intermediate Pareto.] For all $S\in \mathcal{B} $ and $x\in F(S)$, there is no $y \in S$ such that $y \gg x$. Furthermore, for some $x' \in F(S)$, there is no $y' \in S$ such that $y' > x'$. 
\end{description}

The next axiom is about the invariance property under player-wise positive scalar multiplication. This property can be interpreted as postulating that the chosen outcomes are determined in proportion to the utility levels that the players can achieve.
The formal definition is as follows: 
\begin{description}
    \item[\bf Scale Invariance.] For all $a \in \mathbb{R}^n_{++}$, $F(a S) = a F(S)$.
\end{description}

We then introduce an axiom of impartiality among the players. 
\begin{description}
    \item[\bf Anonymity.] For all symmetric problems $S\in \mathcal{B}$, if $x\in F(S)$, then $ x^\pi \in F(S)$ for all  $\pi \in \Pi$. 
\end{description}

Axioms about continuity have also been widely considered.
The following axiom requires that small changes in bargaining situations do not lead to large changes in the chosen outcomes.
\begin{description} 
    \item[\bf Continuity.] For all $S\in \mathcal{B}$ and $x\in S$, if there exists $\{S^k \}_{k\in \mathbb{N}} \subset \mathcal{B}$ and $\{ x^k\}_{k\in \mathbb{N}}\subset \mathbb{R}^n$ such that (i)  $x^k \in F(S^k)$ for all $k\in \mathbb{N}$; (ii) $\{S^k \}_{k\in \mathbb{N}}$ converges to $S$ in the Hausdorff topology; (iii) $\{ x^k\}_{k\in \mathbb{N}}$ converges to $x$, then $x\in F(S)$.
\end{description}
These axioms are standard in the literature, and many solutions, including the Nash solution, satisfy them.

We then consider Nash's independence of irrelevant alternatives (IIA). The following is its set-valued version introduced by \citet{arrow1959rational}. 
\begin{description}
    \item[\bf Independence of Irrelevant Alternatives (IIA).] For all $S, S'\in \mathcal{B}$ with $S'\subset S$, if $S' \cap F(S) \neq \emptyset$, then $F(S') = S' \cap F(S)$. 
\end{description}
Nash's original IIA postulates that if a solution outcome in the original problem is feasible in a smaller problem, then it should also be chosen in the smaller problem.
One of the reasons why Nash's IIA has been criticized is that it requires the consistency property even when the balance of bargaining powers changes significantly. 
Our axiom imposes this contraction independence only when  the maximum utility levels that  players can achieve are equal in both the original problem and the contracted problem. 
\begin{description}
    \item[\bf Weak Independence of Irrelevant Alternatives (Weak IIA).] For all $S, S'\in \mathcal{B}^\text{Eq}$ with $S'\subset S$, if $S' \cap F(S) \neq \emptyset$, then $F(S') = S' \cap F(S)$. 
\end{description}
Similar weak axioms have been introduced, for example, by \citet{DUBRA2001131} and \citet{xu2006alternative}. These axioms were used to characterize the Kalai-Smorodinsky solution or its variants.

\section{A general representation theorem}
\label{sec2_main}

By imposing the axioms introduced in the previous section, we derive a new class of solutions. Before providing our result, we introduce additional definitions: Let $\Delta$ be the set of weights over the players, i.e., $\Delta = \{ w\in \mathbb{R}_+^n \mid \sum_{i\in N} w_i = 1\} $. 
We say that a function $c : \Delta \rightarrow \mathbb{R}_{++} \cup \{  + \infty \}$ is \textit{log-convex} if the function $\log (c (\cdot))$ from $\{ w \in \Delta \mid c(w) < + \infty  \}$ to $\mathbb{R}$ is a convex function. 
We say that a collection $\mathbb{C}$ of functions $c: \Delta \rightarrow \mathbb{R}_{++}\cup \{  + \infty \}$ is \textit{symmetric} if for all $c\in \mathbb{C}$ and $\pi\in \Pi$, the function $c^\pi$ defined as for all $w\in \Delta$, $c^\pi (w) = c(w^\pi)$ is also in $\mathbb{C}$. 

\citet{nakamura2025wp} characterized the solutions satisfying all of the axioms introduced in the previous section. 
Nakamura showed that these solutions can be rationalized by a homogeneous, symmetric, continuous function on utility vectors normalized by the maximal utility levels that the players can achieve. 
Now, we provide an alternative way of representing these solutions. 
The following result reveals that these solutions can be written as generalizations of the Nash solution. 
The key difference is that, in the solutions we characterize, the weight over the players is not fixed but are determined endogenously.

\begin{theorem}
\label{thm_main}
    A solution $F$ satisfies \textit{intermediate Pareto}, \textit{scale invariance}, \textit{anonymity}, \textit{continuity}, and \textit{weak IIA}
    if and only if
    there exists a nonempty symmetric collections $\mathbb{C}$ of log-convex functions $c : \Delta \rightarrow \mathbb{R}_{++}\cup \{  + \infty \} $ such that
    $\max_{c\in \mathbb{C}} \min_{w\in\Delta} c(w) = 1$ and  for all $S\in \mathcal{B}$, 
    \begin{equation}
    \label{eq_dualNKS}
        F (S) = \argmax_{x\in S} \qty{ \max_{c\in \mathbb{C}} \min_{w\in \Delta } c(w) \prod_{i\in N} \qty( {x_i \over b_i (S)} )^{w_i} }. 
    \end{equation}
\end{theorem}

We then discuss the interpretation of \eqref{eq_dualNKS}. This solution  can be understood as a choice rule of an optimistic social planner who consults inequality-averse experts, each of whom is labeled by $c\in \mathbb{C}$. Given a problem $S$, the planner evaluates each outcome $x \in S$ as follows:
\begin{enumerate}
    \item All experts normalize utility levels by $b(S)$, that is, the outcome $x$ is evaluated based on the extent to which the maximum achievable utility value is fulfilled.
    
    \item For each $w\in \Delta$, the expert labeled by $c$ calculates the weighted product of the normalized utility levels and then  multiplies  it by $c(w)$. The value $c(w)$ represents the expert's confidence level in $w$, similar to the confidence expected utility model proposed by \citet{chateauneuf2009ambiguity}. 
    
    \item The expert  labeled by $c$ evaluates $x$  in an inequality-averse way using
    \begin{equation*}
       \min_{w\in \Delta} c(w) \prod_{i\in N} \qty({x_i\over b_i (S)})^{w_i} 
    \end{equation*}
    and suggests this value to the planner. 
    Note that as $c(w)$ decreases, the weight $w$ becomes  less likely to be used. 
    That is, the smaller $c(w)$ is, the more reasonable expert thinks it  to be. 
    \item Finally, the optimistic planner chooses the maximum value among the evaluation values suggested by the experts. 
\end{enumerate}

The Nash solution is a special case of \eqref{eq_dualNKS}. 
It is easy to verify that when $\mathbb{C} = \{ c^\text{Nash} \}$ where $c^\text{Nash} (1/n, 1/n, \cdots, 1/n ) = 1 $ and $c^\text{Nash} (w) = + \infty$ for all $w \neq  (1/n, 1/n, \cdots, 1/n ) $, the solution \eqref{eq_dualNKS} coincides with the Nash solution. 
Furthermore, the Kalai-Smorodinsky solution is also in this class. 
When $\mathbb{C} = \{ c^\text{KS} \}$ where $c^\text{KS} (w) = 1 $ for all $w \in \Delta$, the solution \eqref{eq_dualNKS} coincides with the Kalai-Smorodinsky solution. 

Mathematically speaking, this result is closely related to Theorem 3 of \citet{chandrasekher2022dual}. 
In the proof, we first use Theorem 2 of \citet{nakamura2025wp} to obtain a homogeneous function representing the choice rule. Then, by applying some transformation to that function, we derive an additive function  with respect to the constant vectors. (For a formal definition, see Claim \ref{claim_taransinv} in Appendix.) 
Theorem 3 of \citet{chandrasekher2022dual} provides a representation of the functions that satisfy this property along with several basic ones. By using this axiom, we can rewrite the transformed function. 
Finally, we apply the inverse transformation to revert to the original function and obtain the objective function in \eqref{eq_dualNKS}.

Note that the axioms in Theorem \ref{thm_main} are independent. (i) The solution that assigns $\mathbf{0}$ to any problem satisfies all the axioms except for \textit{intermediate Pareto}.  
(ii) The utilitarian solution, which assigns the maximizers of total utility level to each problem, satisfies all the axioms except for \textit{scale invariance}.
(iii) For each $i\in N$, the $i$-dictatorship solution, which assigns the maximizers of $i$'s utility level, satisfies all the axioms except for \textit{anonymity}. 
(iv) Let $\geq_\text{lex}$ be the binary relation over $\mathbb{R}^n$ such that for all $x,y\in\mathbb{R}^n$, 
\begin{align*}
    x >_\text{lex} y &\iff [ ~ \exists j \in  N  ~~~ \text{s.t.} ~~~ x_{(j-1)} = y_{(j-1)} ~~~\text{and} ~~~ x_{(j)} > y_{(j)} ],
    \\
    x =_\text{lex} y &\iff [ ~ x_{(i)} = y_{(i)} ~~~\text{for all $i\in N$} ~ ]. 
\end{align*}
Consider the solution $F$ defined as for all $S\in \mathcal{B}$, 
\begin{equation*}
    F(S) = \qty{ x\in S ~ \bigg| ~ \nexists  y\in S ~~ \text{s.t.} ~~ \qty({y_1\over b_1 (S)}, \cdots, {y_n \over b_n (S)}  )  >_\text{lex}  \qty({x_1\over b_1 (S)}, \cdots, {x_n \over b_n (S)}  )}. 
\end{equation*}
This is a lexicographic version of the Kalai-Smorodinsky solution and  satisfies all the axioms for \textit{continuity}. 
(v) The solution that assigns to each $S\in \mathcal{B}$ the set of weakly Pareto optimal utility vectors satisfies all the axioms for \textit{weak IIA}.


\section{Special cases: bargaining over linear technologies}
\label{sec2_tech}

This section considers a more specific context, bargaining over production technologies.
Suppose that there are several feasible profiles of production technologies, and the arbitrator chooses the most desirable profiles from them. 
Players live for many periods, and in each period, they produce output goods using technologies and goods they own in that period as inputs.
We abstract from consumption, that is, each player can use all of their goods as input in each period. 
We assume that each technology is linear in the sense that the amount of output is proportional to the amount of input. 
Therefore, each technology can be represented by a positive real number and a profile of technologies can be written as a vector in $\mathbb{R}_{++}^n$. 
Let $T\subset \mathbb{R}_{++}^n$ be a set of profiles of technologies, and let
its typical element be denoted by $t = (t_1, t_2,\cdots, t_n) \in \mathbb{R}_{++}^n$. 

Each player $i$ has initial endowments $e_i \in \mathbb{R}_{++}$. They use this as input for production in the first period.  
Given a technology $t$ and a profile $e = (e_1, e_2,\cdots, e_n) \in \mathbb{R}_{++}^n$ of initial endowments, the profile of payoffs at the end of the first period can be represented by $(t_1 e_1, t_2 e_2, \cdots, t_n e_n)$. (Similarly, if the players use the same profile $t$ for two periods, then the profile of payoffs at the end of the second period can be represented by $(t^2_1 e_1, t^2_2 e_2, \cdots, t^2_n e_n)$.)
Therefore, given a set $T \subset \mathbb{R}_{++}^n$ of feasible profiles of technologies and a profile of initial endowments $e$, the feasible set of payoff vectors at the end of the first period is $eT$. 
Assuming that goods are disposable, the feasible set of payoff vectors becomes $\text{cmp}  (eT )$ and it is an element of $\mathcal{B}$. 

We assume that the arbitrator is concerned solely with the distribution of goods; in other words, the arbitrator is indifferent to the production process and its timing.\footnote{Therefore, we assume that the arbitrator considers the case where an individual $i$ produces goods from an input $e_i = 1$ by using technology $t_i = 2$ and the case  where  $i$ produces goods from an input $e_i = 2$ by using technology $t_i = 1$. Furthermore, the arbitrator identifies the case where an individual $j$ produces goods from an input $e_j = 1$ by using technology $t_i = 2$ for two periods and the case  where an individual $i$ produces goods from an input $e_i = 4$ by using technology $t_i = 1$ for two periods. (This is because individual $j$ obtains $4$ units of goods in both cases.)}
Then bargaining solutions can be interpreted as the arbitrator's decision rules for choosing desirable profiles of technologies.

In this model, the axioms in Theorem \ref{thm_main} can be considered reasonable and interpretable. For example, if the arbitrator thinks that the players' initial endowments are determined by the effort levels fully responsible for them, their choice rule should be independent of players' endowments and satisfy \textit{scale invariance}. 
Thus, the solutions characterized in Theorem \ref{thm_main} are also admissible. 
However, the class of solutions remains overly broad. To address this, we introduce axioms that are particularly compelling in this context to further narrow the range of admissible solutions.

\subsection{Independence of Timing}
\label{subsec_DC}

We then introduce axioms for choice rules of the arbitrator of bargaining over linear technologies. 
The first axiom is about dynamic consistency. 
Suppose that the arbitrator chooses a profile of technologies based on the distribution of goods at the end of the first period. 
The next axiom postulates that if players commit their technologies for several periods, then the arbitrator's choice of profiles of technologies should not depend on the timing when the arbitrator evaluates distributions.

To formalize the above property, we introduce additional notation. For $x\in \mathbb{R}^n_+$ and $m\in \mathbb{N}$, let $x^m = (x_1^m, x_2^m, \cdots, x_n^m)$. For $S\in \mathcal{B}$ and $m\in \mathbb{N}$, let $S^m =  \{ x^m \in \mathbb{R}^n_+ \mid x\in S\}$, that is, $S^m$ is the problem generated by the distribution of goods at the end of period $m$ when $e =\mathbf{1}$.\footnote{To formalize the axiom concisely, we focus on the cases where $e = \mathbf{1}$. Note that this prerequisite can be removed by introducing more complicated notations to deal with the case where $e \neq \mathbf{1}$. } 
Then, our axiom is formalized as follows: 

\begin{description}
    \item[\bf Independence of Timing.] For all $S\in \mathcal{B}$, $x\in S$, and $m\in \mathbb{N}$, $x\in F(S)$ if and only if $x^m \in F(S^m)$. 
\end{description}

By adding this axiom, we can pin down a more specific class of solutions: The structure of the collection $\mathbb{C}$ in Theorem \ref{thm_main} can be restricted, and each function in $\mathbb{C}$ takes only two values, $1$ or $+\infty$. As a result, the objective function can be represented using a collection of subsets of weights, instead of a collection of functions.
The objective function over normalized utility vectors can still be formulated as a two-stage optimization problem, which determines an ``optimal" weight to evaluate each outcome. 

We say that a collection $\mathcal{W}$ of subsets of $\Delta$ is \textit{symmetric} if for all $W\in\mathcal{W}$ and $\pi \in \Pi$, $\{x^\pi \mid x\in W\} \in \mathcal{W}$. We assume that the collection of subsets of $\Delta$ is endowed with the Hausdorff  topology. 
Then, our result is as follows: 

\begin{proposition}
\label{prop_dualself}
    A solution $F$ satisfies \textit{intermediate Pareto}, \textit{scale invariance}, \textit{anonymity}, \textit{continuity}, \textit{weak IIA}, and \textit{independence of timing}
    if and only if
    there exists a nonempty, symmetric, compact collection $\mathcal{W}$ of nonempty, closed, convex subsets of $\Delta$ such that  for all $S\in \mathcal{B}$, 
    \begin{equation}
    \label{eq_dualNKS2}
        F (S) = \argmax_{x\in S} \qty{ \max_{W\in \mathcal{W}} \min_{w\in W }  \prod_{i\in N} \qty( {x_i \over b_i (S)} )^{w_i} }. 
    \end{equation}
\end{proposition}

Compared with Theorem \ref{thm_main}, the above representation can be interpreted as each expert having a set $W\in \mathcal{W}$ of reasonable weights over players, instead of a function $c$ representing the confidence level of each weight. 
While Theorem \ref{thm_main} is obtained using Theorem 2 of \citet{chandrasekher2022dual}, Proposition \ref{prop_dualself} can be proved using Theorem 1 of the same paper.

\subsection{Combination improvement}
\label{subsec_CI}

Next, we consider an axiom related to fairness. Consider a problem $\text{cmp} (eT) \in \mathcal{B}^\text{Eq}$, where $e = \mathbf{1}$, and suppose that $x, y\in F(\text{cmp} (eT))$.\footnote{Note that as in the previous section, we consider the cases where $e = \mathbf{1}$ only to formalize the axiom concisely.} 
Then these two technologies are not necessarily fair outcomes: They may be outcomes chosen by the arbitrator since there is no both efficient and equitable technology. 
The next axiom requires that in these situations,  using technology $x$ in the first period and then switching to technology $y$ in the second period is weakly preferable to committing to $x$ or $y$.
By switching from $x$ to $y$ (or from $y$ to $x$), the players can achieve a more moderate distribution. 

Again, we introduce additional notation. 
For $S \in \mathcal{B}$, let $S^\ast = \{ xy \in \mathbb{R}^n_+ \mid x, y\in S \}$, that is,  $S^\ast$  represents the two-period problem generated from the one-period problem $S = \text{cmp} (eT)$ with $e = \mathbf{1}$. 
In $S^\ast$, the players can use different profiles of technologies in the first and second periods, unlike in $S^2$. 
Then, our axiom of fairness is formalized as follows: 

\begin{description}
    \item[\bf Combination Improvement.] For all $S\in \mathcal{B}^\text{Eq}$ and $x, y \in F(S)$, there exists $z \geq xy $ such that  $z \in F(S^\ast)$. 
\end{description}

If we impose the above axiom together with the axioms in Proposition \ref{prop_dualself}, then the collection $\mathcal{W}$ becomes a singleton. 
That is, these solutions can be interpreted as choice rules of an arbitrator who consults one inequality-averse expert, or as choice rules of a benevolent arbitrator evaluating each outcome independently. 

\begin{proposition}
\label{prop_maxmin}
    A solution $F$ satisfies \textit{intermediate Pareto}, \textit{scale invariance}, \textit{anonymity}, \textit{continuity}, \textit{weak IIA}, \textit{independence of timing}, and \textit{combination improvement}
    if and only if
    there exists a nonempty, closed, convex subset $W$ of $\Delta$ such that  for all $S\in \mathcal{B}$, 
    \begin{equation}
    \label{eq_dualNKS3}
        F (S) = \argmax_{x\in S} \qty{ \min_{w\in W }  \prod_{i\in N} \qty( {x_i \over b_i (S)} )^{w_i} }. 
    \end{equation}
\end{proposition}

This is the counterpart of the maxmin expected utility model (\citealp{gilboa1989maxmin}) to the Nash solution. The solution \eqref{eq_dualNKS3} evaluates each outcome by the minimum weighted product of the normalized utility levels among $W$. 
Note that these solutions still include the Nash solution and the Kalai-Smorodinsky solution as special cases:  
When $W = \{ (1/n, 1/n, \cdots, 1/n ) \}$, the solution \eqref{eq_dualNKS3} is the Nash solution; when $W = \Delta$, the solution \eqref{eq_dualNKS3} is the Kalai-Smorodinsky  solution.

\section{Full IIA and homogeneity}
\label{sec2_homo}

We have considered bargaining solutions that satisfy \textit{weak IIA} and the standard axioms. 
\textit{Weak IIA} and \textit{scale invariance} are compelling when the arbitrator thinks that reflecting the players' bargaining powers is desirable. 
However, if the goal is to maximize a social welfare index defined exogenously, then  the criticism of \textit{IIA} becomes invalid, and rather, \textit{scale invariance} becomes questionable since it postulates that solutions should be responsive to changes in problems. 

Instead of \textit{scale invariance}, consider the following standard condition: 
\begin{description}
    \item[\bf Homogeneity.] For all $\alpha\in \mathbb{R}_{++}$, $F( (\alpha \mathbf{1}) S) = (\alpha \mathbf{1}) F(  S)$
\end{description}

If we replace \textit{weak IIA} and  \textit{scale invariance} with \textit{IIA} and \textit{homogeneity} in Theorem \ref{thm_main}, we can obtain a similar characterization result.  
In this case, the arbitrator directly uses the players' utility levels instead of the normalized ones.

\begin{theorem}
\label{thm_homo}
    A solution $F$ satisfies \textit{intermediate Pareto}, \textit{homogeneity}, \textit{anonymity}, \textit{continuity}, and \textit{IIA}
    if and only if
    there exists a nonempty symmetric collection $\mathbb{C}$ of log-convex functions $c : \Delta \rightarrow \mathbb{R}_{++}\cup \{  + \infty \} $ such that
    $\max_{c\in \mathbb{C}} \min_{w\in\Delta} c(w) = 1$ and  for all $S\in \mathcal{B}$, 
    \begin{equation}
    \label{eq_dualhomo}
        F (S) = \argmax_{x\in S} \qty{ \max_{c\in \mathbb{C}} \min_{w\in \Delta } c(w) \prod_{i\in N} x_i^{w_i} }. 
    \end{equation}
\end{theorem}

Since we can prove this theorem in a similar way to Theorem  \ref{thm_main}, we omit a proof. 
Similar to Theorem \ref{thm_main}, the class of solutions in Theorem \ref{thm_homo} bridges the Nash solution and the egalitarian solution:\footnote{Here, we say that a solution $F$ is the egalitarian solution if $F$ assigns the maximizer of $\min_{i\in N} x_i$ to each problem. 
The definition is slightly different from \citeauthor{Kalai1977Econometrica}'s (1977) original definition, but they coincide in the convex problems. 
}  
When $\mathbb{C} = \{ c^\text{Nash} \}$ where $c^\text{Nash} (1/n, 1/n, \cdots, 1/n ) = 1 $ and $c^\text{Nash} (w) = + \infty$ for all $w \neq  (1/n, 1/n, \cdots, 1/n ) $, the solution in Theorem \ref{thm_homo} is the Nash solution; 
When $\mathbb{C} = \{ c^\text{E} \}$ where $c^\text{E} (w) = 1 $ for all $w \in \Delta$, the solution in Theorem \ref{thm_homo} is the egalitarian solution. 
Furthermore, this class includes the utilitarian solution:\footnote{
We say that a solution $F$ is the utilitarian solution if for all $S\in\mathcal{B}$, 
\begin{equation*}
    F(S) = \argmax_{x\in S} \sum_{i\in N} x_i. 
\end{equation*}
} It is easy to verify that the utilitarian solution satisfies all the axioms.

The axioms in Theorem \ref{thm_homo} are independent. Counterexamples can be constructed in a manner similar to those in Theorem \ref{thm_main}. 

\section{Concluding remark}
\label{sec2_concl}

Motivated by the criticisms of \textit{IIA} introduced by \citet{nash1950bargaining}, we have examined a weak version of it.  
Our main result shows that even if \textit{weak IIA} is imposed instead of the original one, the solutions can still be represented as having a similar structure to the Nash solution (Theorem \ref{thm_main}). 
However, this class of solutions is too broad, so we have characterized more specific classes by considering a concrete context: bargaining over production technologies (Proposition \ref{prop_dualself} and \ref{prop_maxmin}). 
Furthermore, we have shown that the structure obtained in Theorem \ref{thm_main} can also be derived by dropping \textit{scale invariance} instead of \textit{IIA}. 

The classes of solutions characterized in this paper include the Nash solution and the Kalai-Smorodinsky solution (Theorem \ref{thm_main} and Proposition \ref{prop_dualself} and \ref{prop_maxmin}), or the Nash solution, the egalitarian solution, and the utilitarian solution (Theorem \ref{thm_homo}). 
The relationship among these three solutions has been considered in \citet{cao1982preference} and  a series of studies by 
\citet{rachmilevitch2015nash,rachmilevitch2016egalitarian,rachmilevitch2023nash,rachmilevitch2024nash}. Examining the differences among these solutions at the level of parameters in our representations would provide new insights into the relationships among them.



\appendix
\section{Appendix: proofs of the results}

In each proof, we only prove the only-if part. 

Before providing proofs, we introduce additional definitions. Let $D$ be a nonempty  subset of $\mathbb{R}^n$. 
We say that a function $V:D\to \mathbb{R}$ is monotone if for all $x,y\in D$ such that $x\gg y$, $V(x) > V(y)$; A function $V:D\to \mathbb{R}$ is symmetric if for all $x\in D$ and $\pi \in \Pi$, $V(x) = V(x^\pi)$; A function $V:D\to \mathbb{R}$ is homogeneous if for all  $x\in D$ and $\alpha > 0$, $V( \alpha x) = \alpha V(x)$. 

\subsection{Proof of Theorem 1}

Let $F$ be a bargaining solution that satisfies  \textit{intermediate Pareto}, \textit{scale invariance}, \textit{anonymity}, \textit{continuity}, and \textit{weak IIA}. 

\citet{nakamura2025wp} showed the following result:
\begin{lem}[Theorem 2 in \citet{nakamura2025wp}]
\label{lem_normrational}
    A solution $F$ satisfies \textit{intermediate Pareto}, \textit{scale invariance}, \textit{anonymity}, \textit{continuity}, and \textit{weak IIA}
    if and only if there exists a monotone, symmetric, homogeneous, and continuous function $V:\mathbb{R}^n_+ \rightarrow \mathbb{R}$ such that for all $S\in \mathcal{B}$, 
    \begin{equation}
    \label{eq_homogrepre}
        F (S) = \argmax_{x\in S}  ~ V \qty( {x_1 \over b_1 (S) }, {x_2 \over b_2 (S) }, \cdots, {x_n \over b_n (S) }). 
    \end{equation}
\end{lem}
\vspace{3mm}

By the argument in \citet{nakamura2025wp}, the function $V$ in \eqref{eq_homogrepre} is defined as for all $x\in \mathbb{R}^n_+$, 
    \begin{equation}
    \label{eq_defV}
        V(x) = \inf \{ \alpha \in \mathbb{R}_+ \mid\alpha > 0 ~~\text{and} ~~ \alpha \mathbf{1} \in F( \text{scmp} \{ x, \alpha \mathbf{1} \}  ) \}. 
    \end{equation}
By the definition,  $V(\alpha \mathbf{1}) = \alpha$  for all $\alpha \in \mathbb{R}_+$. 
By Lemma \ref{lem_normrational}, since $V$ is homogeneous, for all $S\in \mathcal{B}^\text{Eq}$ with $b(S) = \mathbf{1}$, 
\begin{equation}
\label{eq_Be_rational}
        F (S) = \argmax_{x\in S} ~ V (x_1, x_2, \cdots, x_n). 
\end{equation}

Note that for all $x\in\mathbb{R}^n_{++}$, $V(x) > 0$. Indeed, for $\alpha \in  (0, x_{(1)} )$, \textit{intermediate Pareto} implies that $\alpha \mathbf{1} \notin F( \text{scmp} \{ x, \alpha \mathbf{1} \}  ) $. By \eqref{eq_defV}, we have $V(x) > V(\alpha \mathbf{1}) = \alpha > 0$. 


Now we define the function $I: \mathbb{R}^n\rightarrow \mathbb{R}$ as for all $x\in\mathbb{R}^n$,
\begin{equation}
\label{eq_defI}
    I(x) = \log V(e^{x_1}, e^{x_2}, \cdots, e^{x_n}). 
\end{equation}
Since $(e^{x_1}, e^{x_2}, \cdots, e^{x_n}) \gg \mathbf{0}$ for all $x\in \mathbb{R}^n$, we have $ V(e^{x_1}, e^{x_2}, \cdots, e^{x_n}) > 0$, which implies that $I$ can be defined over $\mathbb{R}^n$. 

\begin{claim}
    The function $I$ is monotone, symmetric, and continuous. 
\end{claim}

\begin{proof}
We only verify that $I$ is monotone. 
Let $x,y\in\mathbb{R}^n$ with $x\gg y$. Since $(e^{x_1}, e^{x_2}, \cdots, e^{x_n}) \gg (e^{y_1}, e^{y_2}, \cdots, e^{y_n})$ and $V$ is monotone, we have $V(e^{x_1}, e^{x_2}, \cdots, e^{x_n}) > V(e^{y_1}, e^{y_2}, \cdots, e^{y_n})$. 
Therefore, $I(x) = \log V(e^{x_1}, e^{x_2}, \cdots, e^{x_n}) > \log V(e^{y_1}, e^{y_2}, \cdots, e^{y_n}) = I(y)$. 
The other properties also follow  from the counter properties of $V$. 
\end{proof}

\begin{claim}
\label{claim_taransinv}
    For all $x\in \mathbb{R}^n$ and $\alpha \in \mathbb{R}$, $I(x + \alpha \mathbf{1}) = I(x) + \alpha$. 
\end{claim}

\begin{proof}
If $\alpha = 0$, then the proof is trivial. Let $x\in \mathbb{R}^n$ and $\alpha \in \mathbb{R}_+$. Since $V$ is homogeneous,  we have 
\begin{align*}
    I(x + \alpha \mathbf{1}) 
    &= \log V(e^{x_1 + \alpha}, e^{x_2+ \alpha}, \cdots, e^{x_n + \alpha}) \\
    &= \log e^\alpha V(e^{x_1}, e^{x_2}, \cdots, e^{x_n}) \\
    &= I(x) + \alpha. 
\end{align*}
If $\alpha < 0$, then by applying the result of the previous case, we have $I(x) = I(x + \alpha \mathbf{1} - \alpha \mathbf{1}) =I(x + \alpha \mathbf{1}) - \alpha$, that is, $I(x + \alpha \mathbf{1}) = I(x) + \alpha$. 
\end{proof}

Then we can apply the argument of Theorem 3 in  \citet{chandrasekher2022dual}. 
There exists a nonempty collection $\Phi$ of convex functions $\varphi: \Delta \rightarrow \mathbb{R} \cup \{ + \infty \}$ such that  $\max_{\varphi\in \Phi} \min_{w\in\Delta} \varphi(w) = 0$ and for all $x\in \mathbb{R}^n$, 
\begin{equation*}
    I(x) = \max_{\varphi\in \Phi} \min_{w\in \Delta} \qty{ \sum_{i\in N} w_i x_i + \varphi (w) }. 
\end{equation*}
By construction of $\Phi$ in their proof (\citealp{chandrasekher2022OAdual}), Claim 1 implies that we can set $\Phi$ to be symmetric.

By \eqref{eq_defI}, for all $y\in \mathbb{R}^n$, 
    \begin{equation*}
        V(e^{y_1}, e^{y_2}, \cdots, e^{y_n}) = e^{I(y)}. 
    \end{equation*}
    Therefore, for all $x\in \mathbb{R}^n_{++}$, 
    \begin{align*}
        V(x) 
        &= \text{exp} \big(  I(\log x_1, \log x_2, \cdots, \log x_n) \big)\\
        &= \text{exp} \qty( \max_{\varphi \in \Phi} \min_{w\in \Delta} \qty{ \sum_{i\in N} w_i \log x_i + \varphi (w) } )\\
        &= \max_{\varphi \in \Phi} \min_{w\in \Delta}  \text{exp} \qty(\sum_{i\in N} w_i \log x_i + \varphi (w) ) \\
        &= \max_{\varphi \in \Phi} \min_{w\in \Delta}  \text{exp} \big( \varphi (w) \big) \prod_{i\in N} \text{exp} \big( w_i \log x_i  \big). 
    \end{align*}
For each $\varphi \in \Phi$,  define the function $c_\varphi:\Delta \rightarrow \mathbb{R}\cup\{ +\infty \}$ as for all $w\in\Delta$, $c_\varphi (w) =\text{exp} \big( \varphi (w) \big) $. 
Let $\mathbb{C} = \{ c_\varphi\}_{\varphi\in \Phi}$.  
    Then, for all $x\in (0,1]^n$, 
    \begin{equation*}
        V(x) = \max_{c\in \mathbb{C}} \min_{w\in \Delta } c(w) \prod_{i\in N} x_i^{w_i}.
    \end{equation*}
    By the continuity of $V$, it can be extended to $[0,1]^n$. 

    Therefore, for all $S\in \mathcal{B}$, 
    \begin{equation*}
        F (S) = \argmax_{x\in S} \qty{ \max_{c\in \mathbb{C}} \min_{w\in \Delta } c(w) \prod_{i\in N} \qty( {x_i \over b_i (S)} )^{w_i} }. 
    \end{equation*}

    Note that by construction, each $c \in \mathbb{C}$ is a symmetric and log-convex function. Furthermore, since $\max_{\varphi\in \Phi} \min_{w\in\Delta} \varphi(w) = 0$, 
    \begin{equation*}
        \max_{\varphi\in \Phi} \min_{w\in\Delta} c(w) = \max_{\varphi\in \Phi} \min_{w\in\Delta} \text{exp} (\varphi (w)) = \text{exp} \qty(\max_{\varphi\in \Phi} \min_{w\in\Delta}  \varphi (w))= 1.
    \end{equation*}

\subsection{Proof of Proposition 1}

Let $F$ be a bargaining solution that satisfies  \textit{intermediate Pareto}, \textit{scale invariance}, \textit{anonymity}, \textit{continuity}, \textit{weak IIA}, and \textit{independence of timing}. 

\begin{claim}
\label{claim_homo}
    For all $x\in \mathbb{R}^n_-$ and $\alpha \in \mathbb{R}_+$, $I(\alpha x) = \alpha I(x)$. 
\end{claim}

\begin{proof}
    By the definition of $V$ and \textit{independence of timing}, for all $y, z\in [0,1]^n$, 
    \begin{equation}
    \label{eq_expindV}
        V(y)\geq  V(z) \iff V(y^m) \geq V(z^m).
    \end{equation}
    By \eqref{eq_expindV} and the definition of $I$, for all $y',z' \in \mathbb{R}_-^n$, 
    \begin{equation}
    \label{eq_expind}
        I(y' )\geq  I(z') \iff I(m y') \geq I(m z'). 
    \end{equation}

    We prove that for all $x\in \mathbb{R}^n_-$ and $\alpha \in \mathbb{R}_+$, $I(\alpha x) = \alpha I(x)$.  
    By the definition of $I$, $I(x) = I (I(x)\mathbf{1})$. 
    By \eqref{eq_expind}, for all $\ell \in \mathbb{N}$, $I(\ell x) = I (\ell I(x)\mathbf{1}) = \ell I(x)$. 
    Similarly, for all $m \in \mathbb{N}$, 
    $I(x/m ) = I ( (I(x)/m) \mathbf{1} ) = I(x)/m$. 
    Therefore, for all $q\in \mathbb{Q}_{++}$, $I(q x) = q I(x)$. 
    By the continuity of $I$, for all $\alpha \in \mathbb{R}_{+}$, $I(\alpha x) = \alpha I(x)$. 
\end{proof}

Then we can apply the argument of Theorem 1 in \citet{chandrasekher2022dual}. 
There exists a nonempty compact collection $\mathcal{W}$ of nonempty, closed, convex subsets of $\Delta$ such that for all $x\in \mathbb{R}_-^n$, 
\begin{equation*}
    I(x) = \max_{W \in \mathcal{W}} \min_{w\in W} \sum_{i\in N} w_i x_i . 
\end{equation*}
By construction of $\mathcal{W}$ in their proof (\citealp{chandrasekher2022OAdual}), Claim 1 implies that we can set $\mathcal{W}$ to be symmetric. 

Therefore, for all $x\in (0,1]^n$, 
\begin{align*}
    V(x) 
    &=  \text{exp} \big(  I(\log x_1, \log x_2, \cdots, \log x_n) \big) \\
    &= \text{exp} \qty(  \max_{W \in \mathcal{W}} \min_{w\in W} \sum_{i\in N} w_i \log x_i ) \\
    &= \max_{W \in \mathcal{W}} \min_{w\in W}  \text{exp} \qty(\sum_{i\in N} w_i \log x_i ) \\
    &= \max_{W \in \mathcal{W}} \min_{w\in W} \prod_{i\in N}  x_i^{w_i} . 
\end{align*}
By the continuity of $V$, it can be extended to $[0,1]^n$. 

Hence, for all $S\in \mathcal{B}$, 
\begin{equation*}
        F (S) = \argmax_{x\in S} \qty{ \max_{W\in \mathcal{W}} \min_{w\in W }  \prod_{i\in N} \qty( {x_i \over b_i (S)} )^{w_i} }. 
    \end{equation*}

\subsection{Proof of Proposition 2}

Let $F$ be a bargaining solution that satisfies  \textit{intermediate Pareto}, \textit{scale invariance}, \textit{anonymity}, \textit{continuity}, \textit{weak IIA}, \textit{independence of timing}, and \textit{combination improvement}. 

\begin{claim}
\label{claim_halfconcave}
    For all $x, y\in \mathbb{R}^n_-$ such that $I(x) = I(y)$, $I(x/2 + y/2) \geq I(x)$. 
\end{claim}

\begin{proof}
    Let $x', y'\in (0,1]^n$ with $V(x') = V(y') (> 0)$, and  let $S = \text{cmp} \{ x', y', e^1, e^2, \cdots, e^n \} \in \mathcal{B}^\text{Eq}$, where, for each $i\in N$, $e^i$ is a vector in $\mathbb{R}^n$ such that $e^i_i = 1$ and $e^i_j = 0$ for all $j\in N \backslash \{ i \}$. 
    Note that by Proposition \ref{prop_dualself}, $V(e^i) = 0$ for all $i\in N$. 
    By Theorem \ref{thm_main}, $x', y'\in F(S)$. 
    By \eqref{eq_expindV}, we have $V(x'^2) = V(y'^2)$. 
    It is straightforward to prove that there is no $z\in S^\ast$ such that $z\geq x'y'$ and  $z\neq x'y'$. 
    By \textit{combination improvement}, we have $x'y' \in F(S^\ast)$. Since $b(S^\ast) = \mathbf{1}$, $V(x'y') \geq V(x'^2) = V(y'^2)$. 
    Therefore, by the definition of $I$, for all $x, y \in \mathbb{R}^n_-$ such that $I(x) = I(y)$, 
    $I(x/2 + y/2)\geq I(x)$. 
\end{proof}

\begin{claim}
    The function $I$ is concave on $\mathbb{R}^n_-$. 
\end{claim}

\begin{proof}
    Let $x', y' \in \mathbb{R}^n_-$. 
    Without loss of generality, assume $I(x') \geq I(y')$. Let $\beta =I(x')  - I(y')$. 
    By Claim \ref{claim_taransinv}, $ I(x'-\beta \mathbf{1}) = I(x') - \beta = I(x') -I(x') +I(y') = I(y')$.
    By Claim \ref{claim_taransinv} and \ref{claim_halfconcave}, 
    \begin{align*}
        I(y') 
        &\leq 
        I\qty({1\over 2} (x'-\beta \mathbf{1}) + {1\over 2} y') \\
        &= 
        I\qty({1\over 2} x'  +  {1\over 2} y') -{1\over 2} \beta \\
        &=  
        I\qty({1\over 2} x'  +  {1\over 2} y') -{1\over 2} I(x') + {1\over 2} I(y'),
    \end{align*}
    which can be rewritten as  
    \begin{equation}
    \label{eq_halfconv}
         I\qty({1\over 2} x' + {1\over 2}  y')
         \geq {1\over 2} I(x') +  {1\over 2} I(y'). 
    \end{equation}

    For all $x,y \in \mathbb{R}^n_-$ and $\alpha \in [0,1]$,  Claim \ref{claim_homo} and \eqref{eq_halfconv} imply that 
    \begin{align*}
        I(\alpha x + (1-\alpha) y )  &=  I \qty( {1\over 2}2\alpha x + {1\over 2} 2(1-\alpha) y )
        \\
        &\geq {1\over 2} I(2\alpha x) + {1\over 2} I(2(1-\alpha) y )\\
        &=  {2\alpha \over 2} I( x) + {2(1-\alpha) \over 2} I( y ) \\
        &= \alpha I (x) + (1 - \alpha ) I (y ). \qedhere
    \end{align*}
\end{proof}

Then we can applying the argument in the proof of Theorem 1 in \citet{gilboa1989maxmin}. 
There exists a nonempty closed convex subset $W$ of $\Delta$  such that for all $x\in \mathbb{R}^n_-$, 
\begin{equation*}
    I(x) =  \min_{w\in W} \sum_{i\in N} w_i x_i . 
\end{equation*}
By the construction of $W$ in their proof and Claim 1, $W$ is symmetric. 

Therefore, for all $x\in (0,1]^n$, 
\begin{align*}
    V(x) 
    &=  \text{exp} \big(  I(\log x_1, \log x_2, \cdots, \log x_n) \big) \\
    &= \text{exp} \qty( \min_{w\in W} \sum_{i\in N} w_i \log x_i ) \\
    &=  \min_{w\in W}  \text{exp} \qty(\sum_{i\in N} w_i \log x_i ) \\
    &= \min_{w\in W} \prod_{i\in N}  x_i^{w_i} . 
\end{align*}
By the continuity of $V$, it can be extended to $[0,1]^n$. 

Hence, for all $S\in \mathcal{B}$, 
\begin{equation*}
        F (S) = \argmax_{x\in S} \qty{  \min_{w\in W }  \prod_{i\in N} \qty( {x_i \over b_i (S)} )^{w_i} }. 
   \end{equation*}


\bibliographystyle{econ}
\bibliography{reference}


\end{document}